%% file: main.tex
\documentclass[sigconf,letterpaper,natbib=false]{acmart}

\usepackage{pdfpages}
\usepackage{tcolorbox}
\usepackage{enumitem}

\newcommand{\roundframe}[1]{{\setlength\fboxrule{0pt}\fbox{\tcbox[colframe=black,colback=white,shrink tight,boxrule=0.5pt,extrude by=2.5pt]{\small #1}}}}

\AtBeginDocument{%
  \providecommand\BibTeX{{%
    \normalfont B\kern-0.5em{\scshape i\kern-0.25em b}\kern-0.8em\TeX}}}

\setcopyright{acmcopyright}
\copyrightyear{2024}
\acmYear{2024}
\acmDOI{10.1145/xxxxxxx.xxxxxxx}

\acmConference[WiSec '24]{WiSec '24: 17th ACM Conference on Security and Privacy in Wireless and Mobile Networks}{May 27--May 30, 2024}{Seoul, Korea}
\acmBooktitle{WiSec '24: 17th ACM Conference on Security and Privacy in Wireless and Mobile Networks, May 27--May 30, 2024, Seoul, Korea}
\acmPrice{15.00}
\acmISBN{978-1-4503-XXXX-X/21/06}

\RequirePackage[
  datamodel=acmdatamodel,
  style=acmnumeric,
  ]{biblatex}
  
\begin{document}

\title{POSTER: Never Gonna Give You Up:\\Exploring Deprecated NULL Ciphers in Commercial VoWiFi Deployments}

\author{Gabriel K. Gegenhuber}
\email{gabriel.gegenhuber@univie.ac.at}
\affiliation{%
  \institution{University of Vienna}
  \department{Faculty of Computer Science}
  \department{Doctoral School Computer Science}
  \city{Vienna}
  \country{Austria}
}
\author{Philipp É. Frenzel}
\email{pfrenzel@sba-research.org}
\affiliation{%
  \institution{SBA Research}
  \city{Vienna}
  \country{Austria}
}
\author{Edgar Weippl}
\email{edgar.weippl@univie.ac.at}
\affiliation{%
  \institution{University of Vienna}
  \department{Faculty of Computer Science}
  \city{Vienna}
  \country{Austria}
}

\renewcommand{\shortauthors}{Gegenhuber et al.}

\begin{abstract}
\input{content/00_abstract}

\end{abstract}



\maketitle

\input{content/01_introduction}

\input{content/02_methodology}

\input{content/03_conclusion}

\printbibliography

\newpage

\includepdf[pages=-]{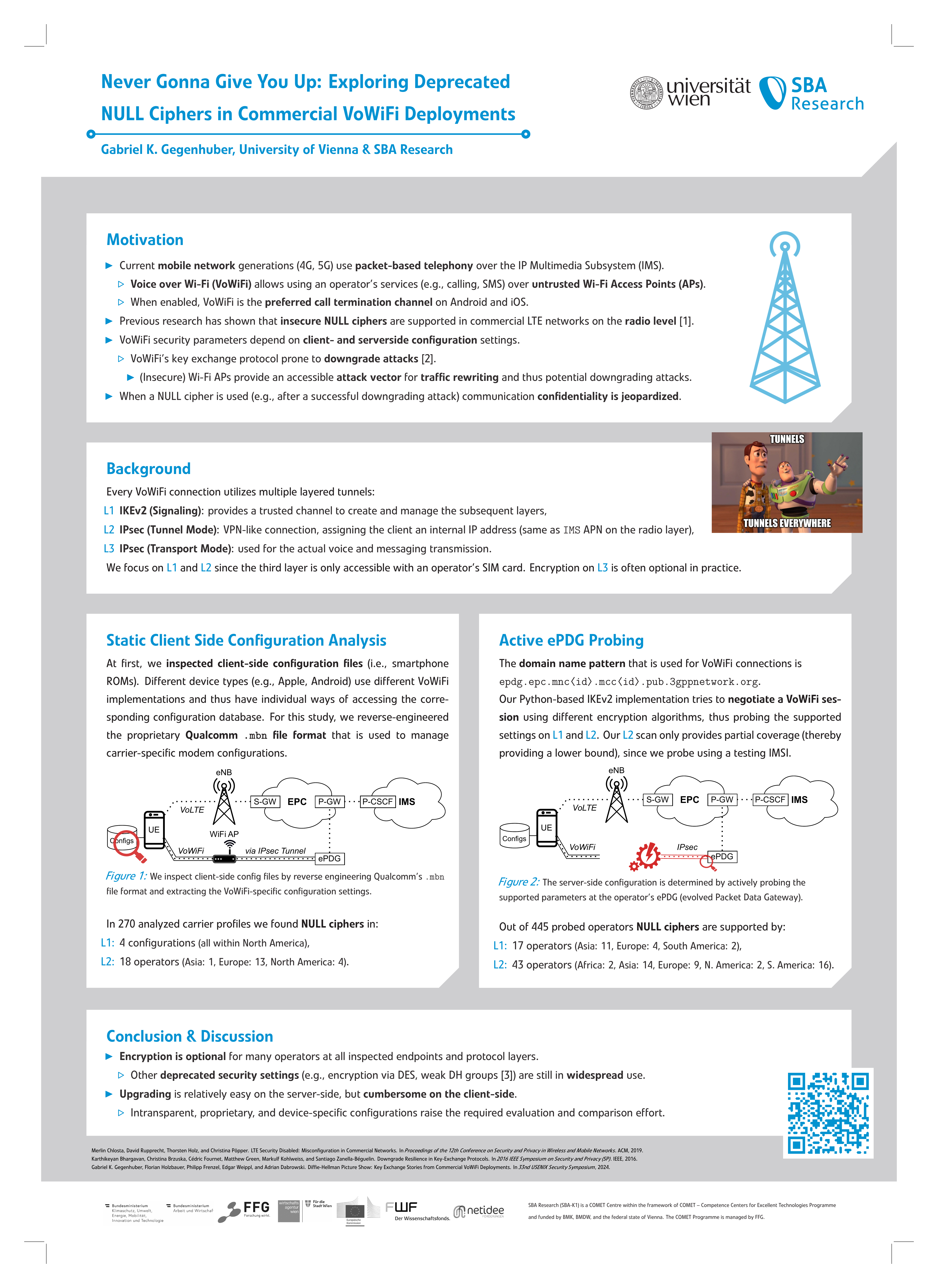}

\end{document}

%% file: content/00_abstract.tex
In today's cellular network evolutions, such as 4G and 5G, the IMS (IP Multimedia Subsystem) serves as a crucial component in managing voice calls and handling short messages.
Besides accessing the IMS over the traditional radio layer, many operators use Voice over Wi-Fi (\mbox{VoWiFi}) allowing customers to dial into their core network over the public Internet using an (insecure) Wi-Fi connection.

To protect against malicious actors on the WiFi or Internet domain, the traffic is sent over a series of IPsec tunnels, ensuring confidentiality and integrity.
Similar to other encrypted protocols (e.g. TLS), the client and server use a handshake protocol (i.e., IKEv2) to communicate their supported security configurations and to agree upon the used parameters (e.g., keys or an encryption algorithm) for the ongoing session.
This however opens the door for security vulnerabilities introduced by misconfiguration.

We want to analyze security configurations within commercial VoWiFi deployments, both on the client and server side, spotting deprecated configurations that undermine communication security.

%% file: content/01_introduction.tex
\section{Introduction}
In today's society, mobile network services play a vital role, with over 5.4 billion people relying on cellular networks for connectivity and communication~\cite{GSMA2023MobileEconomy}.
With 4G currently being the predominant wireless standard and 5G rapidly gaining traction, many operators are actively phasing out older legacy networks (2G and 3G), completing the transition from circuit-switched to a comprehensive packet-switched network paradigm.

In the packet-switched domain, operators utilize VoIP (Voice over IP) based technology to manage voice calls and messages.
While VoLTE (Voice over LTE) uses the traditional radio infrastructure to connect to the operator, VoWiFi (Voice over Wi-Fi) allows customers to use their operator's services over untrusted third-party wireless networks.
Consequently, customers can leverage existing Wi-Fi access points (APs) and continue utilizing their mobile phones for voice calls in areas with poor or no cellular reception.

Beyond that, current operating systems (e.g., Android and iOS) already configure VoWiFi as the preferred call termination channel when enabled.
Due to this integral role in current and upcoming network generations, it is crucial to 
guarantee high security- and privacy standards over VoWiFi.

To establish a trusted communication channel over an untrusted Wi-Fi AP, the User Equipment (UE) client connects to the Evolved Packet Data Gateway (ePDG) server by establishing an IPsec tunnel.
More specifically, they negotiate a Security Association (SA) using the Internet Key Exchange (IKE) protocol.
While the specification recommends using a strong encryption algorithm (e.g., AES), the actual implementation of this recommendation depends on the deployed configuration settings (both on the client- and server-side).

Studies on LTE security configurations at the radio layer have revealed that operators frequently permit insecure settings, such as allowing connections with disabled encryption and integrity algorithms~\cite{chlosta2019lte}.
Additionally, previous work has shown that the used IKEv2 mechanism is prone to downgrade attacks~\cite{bhargavan2016downgrade}, making it crucial to prune insecure settings from the configuration, thus leaving an active attacker (e.g., a malicious Wi-Fi AP operator) no room for downgrading the negotiated SA.

To expose insecure configurations within commercial VoWiFi deployments, we analyze the used configuration settings, by i) inspecting client-side configuration files and ii) probing ePDG servers for supported security parameters.

%% file: content/02_methodology.tex
\section{Methodology}
Large-scale measurement studies focusing on the radio layer and the deployed security configuration practices are scarce due to the decentralized structure of cellular networks, characterized by fragmentation into numerous local operators worldwide.

In contrast to the traditional radio layer that can only be measured locally, VoWiFi exposes parts of an operator's infrastructure to the public Internet, enabling new scanning possibilities for large-scale measurement studies.
In this study, we aim to provide a global overview of insecure and deprecated configuration practices.
For the client-side we do a static analysis of operator-specific files provisioned to real-world VoWiFi clients (i.e., consumer-grade smartphones).
Since configurations for all operators are saved within the firmware ROM, we do not need a SIM card or be geographically close to any operators under inspection.

To also get a picture of the server-side configurations, we actively probe the Internet-accessible ePDG endpoints testing the supported security parameters.
\\\\
Every VoWiFi connection utilizes multiple layered tunnels:
\begin{itemize}
    \item[\roundframe{L1}] \textbf{IKEv2 (Signaling)}: provides a trusted channel to create and manage the subsequent layers
    \item[\roundframe{L2}] \textbf{IPsec (Tunnel Mode)}: VPN-like connection, assigning the client an internal IP address (cf. \texttt{IMS} APN on the radio layer)
    \item[\roundframe{L3}] \textbf{IPsec (Transport Mode)}: used for the actual voice and messaging transmission
\end{itemize}

We focus on the first two layers, since the third layer requires successfully establishment of previous tunnels which is only possible with an operator's SIM card.
Although \roundframe{L3} also supports encryption, it is optional and not required by many operators.

\subsection{Client-Side Configuration Analysis}
Different device types (e.g., Apple, Android) use different VoWiFi implementations and thus have individual ways of saving and accessing the corresponding configuration database.
For this study, we reverse-engineered the proprietary Qualcomm \texttt{.mbn} file format that is used to manage carrier-specific modem configurations (also called \texttt{MCFGs}) on most Qualcomm-based smartphones.
These configuration files can be extracted from the modem image (often named \texttt{NON-HLOS.bin}) that is part of the smartphone ROM.
Using our parsing tool\footnote{\url{https://github.com/sbaresearch/mbn-mcfg-tools}}, we extract the VoWiFi-related settings from a recent smartphone image (i.e., the Xiaomi 13 Pro from Jan. 2024).

\begin{figure}[h]
    \centering
    \includegraphics[page=2,width=1\linewidth]{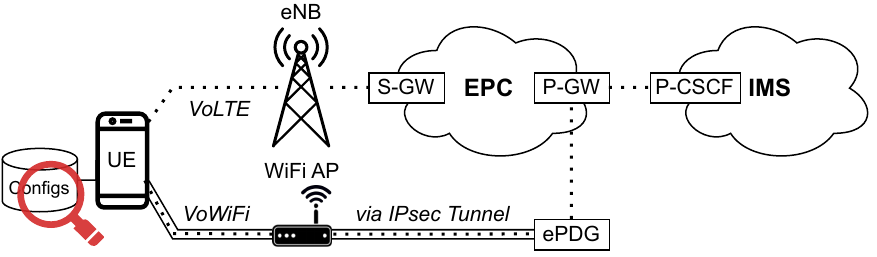}\vspace{-1mm}
    \caption{We inspect client-side config files by reverse engineering Qualcomm's \texttt{.mbn} file format.}%
    \label{fig:VoWifi-overview}%
    \vspace{-2mm}
\end{figure}
Our analyzed configuration database contains 270 different operator-specific client settings.
For \roundframe{L1}, we found four configurations (all within North America) supporting connections without any encryption algorithm.
On the second layer \roundframe{L2}, a total of 18 operators (Asia: 1, Europe: 13, North America: 4) allow unencrypted IPsec connections.
Note that the North American configurations for \roundframe{L1} are also present within the \roundframe{L2} set, thus not requiring encryption on any of the two layers.

\subsection{Active ePDG Probing}
3GPP TS 23.003~\cite{EtsiNumberingAdressingIdentification} specifies the domain name pattern that is used for VoWiFi connections corresponding to an operator's Mobile Country- and Mobile Network Code (MCC, MNC):

{\hspace{0.2cm}\texttt{epdg.epc.mnc\textlangle{}id\textrangle.mcc\textlangle{}id\textrangle.pub.3gppnetwork.org}
\\
Our Python-based IKEv2 implementation tries to negotiate a \mbox{VoWiFi} session using different encryption algorithms, thus probing the supported settings on \roundframe{L1} and \roundframe{L2} for 445 existing domains.
\begin{figure}[h]
    \centering
    \includegraphics[page=4,width=1\linewidth]{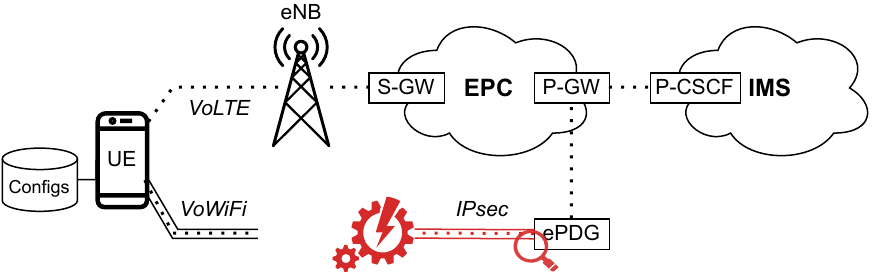}\vspace{-1mm}
    \caption{The server-side config is determined by actively probing the supported parameters at the operator's ePDG.}%
    \label{fig:VoWifi-overview}%
    \vspace{-2mm}
\end{figure}

For \roundframe{L1} a total of 17 ePDG domains
(Asia: 11, Europe: 4, South America: 2)
allow completing the handshake with disabled encryption.
For \roundframe{L2}, at least 43 domains (Africa: 2, Asia: 14, Europe: 9, North America: 2, South America: 16) support unencrypted IPsec connections.
Note that our \roundframe{L2} scan only provides partial coverage, since some operators ignored our requests at this later protocol stage due to not providing a valid IMSI within the operator's range.
Thus, our results provide a lower bound on insecure configurations.

%% file: content/03_conclusion.tex
\section{Conclusion}
We find VoWiFi configurations with optional encryption for all inspected protection layers both on the client- and on the server-side. Apart from the presented findings, other deprecated security settings (e.g., encryption via DES) are still in widespread use.
While removing support for insecure algorithms is relatively easy on the server-side, intransparent, proprietary and device-specific configurations make rolling out changes to existing clients cumbersome.